\documentclass[a4paper]{article}
\usepackage[a4paper,includemp=false,body={16cm,24.7cm},vcentering]{geometry}
\usepackage[utf8]{inputenc}
\usepackage[T1]{fontenc}
\usepackage{mathptmx}
\usepackage[bf,center]{caption}
\usepackage{graphicx}

\usepackage{url,amsmath,amssymb,amsthm}
\usepackage{caption}
\makeatletter
\renewcommand{\section}{\@startsection{section}{1}{0mm}{30pt}{12pt}{\normalfont\normalsize\bfseries}}

\renewcommand{\subsection}{\@startsection{subsection}{2}{0mm}{18pt}{12pt}{\normalfont\normalsize\itshape}}
\makeatother

\newcommand{\Title}[1]{\begin{center}{\bfseries\fontsize{12pt}{12pt}\selectfont#1}\end{center}}
\newcommand{\Author}[2]{\begin{center}{\fontsize{12pt}{12pt}\selectfont#1}\\{\it #2~}\end{center}}
\newcommand{\Introduction}{\section*{Introduction}}

\newcommand{\Acknowledgments}{\section*{Acknowledgements}}

\newcommand{\mechanic}{\texttt{Mechanic}}

\begin{document}

\Title{Mechanic: a new numerical MPI framework for the dynamical astronomy}
  
\Author{
M. S\l{}onina$^1$, K. Go\'zdziewski$^2$, C. Migaszewski$^3$}
{Toru\'n Centre forAstronomy, Nicolaus Copernicus
University, Gagarina 11, 87-100 Toru\'n, Poland \\
$^{1}$slonina@astri.umk.pl, 
$^{2}$k.gozdziewski@astri.umk.pl, 
$^{3}$c.migaszewski@astri.umk.pl}
 
\Introduction

\noindent
In the field of the Solar system and planetary dynamics, extensive 
computational experiments became useful and well established, standard research tools. 
These experiments regard direct numerical integrations of complex equations of motion 
to study the long-term orbital evolution and stability of various $N$-body 
systems (e.g., \cite{laskar2010}), an analysis of the resonance structure of the 
phase space of these systems and particular classes of solutions (like periodic
and quasi-periodic orbits, 
equilibria), calculating dynamical maps \cite{cincotta2003} and 
cross-sections, modeling various types of observations of extrasolar 
planetary systems (radial velocities, astrometric and imaging data, eclipse and TOA timing, 
photometric transits) by quasi-global evolutionary algorithms (e.g.,\cite
{gozdziewski2008}), investigating qualitative features of basic dynamical 
models in the framework of the general dynamical systems theory (e.g., \cite
{science}), spacecraft trajectories optimization (e.g., \cite{conway}), 
just to mention a few subjects of this rapidly developing branch of the 
dynamical astronomy. 

Usually, these experiments are equivalent to performing the same or very 
similar numerical operation or procedure on a large set of initial 
conditions (e.g., numerical integrations, dynamical maps) or intermediate 
data (e.g., when evaluating the objective function during the optimization 
process). In a language of computing, such calculations may be understood 
as standalone numerical tasks taking seconds, but also hours and days of 
single CPU-time. If processing of large sets of such tasks is required, one 
can split a given numerical experiment onto smaller parts, and distribute 
them over a computing pool (usually, CPU cluster or a network of 
workstations). This leads however to task management issues, which may be 
handled efficiently only by dedicated software tools. 

Nowadays, there are different task management systems available. Likely, the 
best example is the well known Condor package \cite{condor}. Within such a 
numerical framework, the user--supplied, stand-alone executable code 
performing computations is distributed over a computing pool. The input and 
output data of each software instance must be then handled by the host 
node. It requires both an efficient, possibly unified task preparation 
scheme, check-pointing and keeping intermediate results, as well as data 
storage and a post--processing (assembling the results from computing 
nodes). Focusing on the dynamical astronomy, we address these issues by 
developing a new management code, called \mechanic{} that is -- unlike 
Condor and similar software systems --  built on the basis of the Message 
Passing Interface (MPI) \cite{mpi}. The MPI is highly-standardized and 
portable message-passing system widely used on a variety of parallel 
computers and CPU-clusters. We shortly present here some key features of 
that software (Sect.~\ref{overview}), and we describe its basic usage for 
investigating a simple dynamical system (Section \ref {arnoldweb}). 
Moreover, \mechanic{} may become a new helper tool in a wide range of 
applications, particularly focusing on processing large data sets. We 
applied it to study the global dynamics of the $\nu$~Octantis planetary 
system, see our second paper in this volume. More details will be described 
elsewhere (Slonina et al., 2012, in preparation).

\section{Overview of the \mechanic{} framework} \label{overview}
\begin{figure}[ht]
\begin{minipage}[t]{0.49\textwidth}
\centering\includegraphics[width=0.9\textwidth]{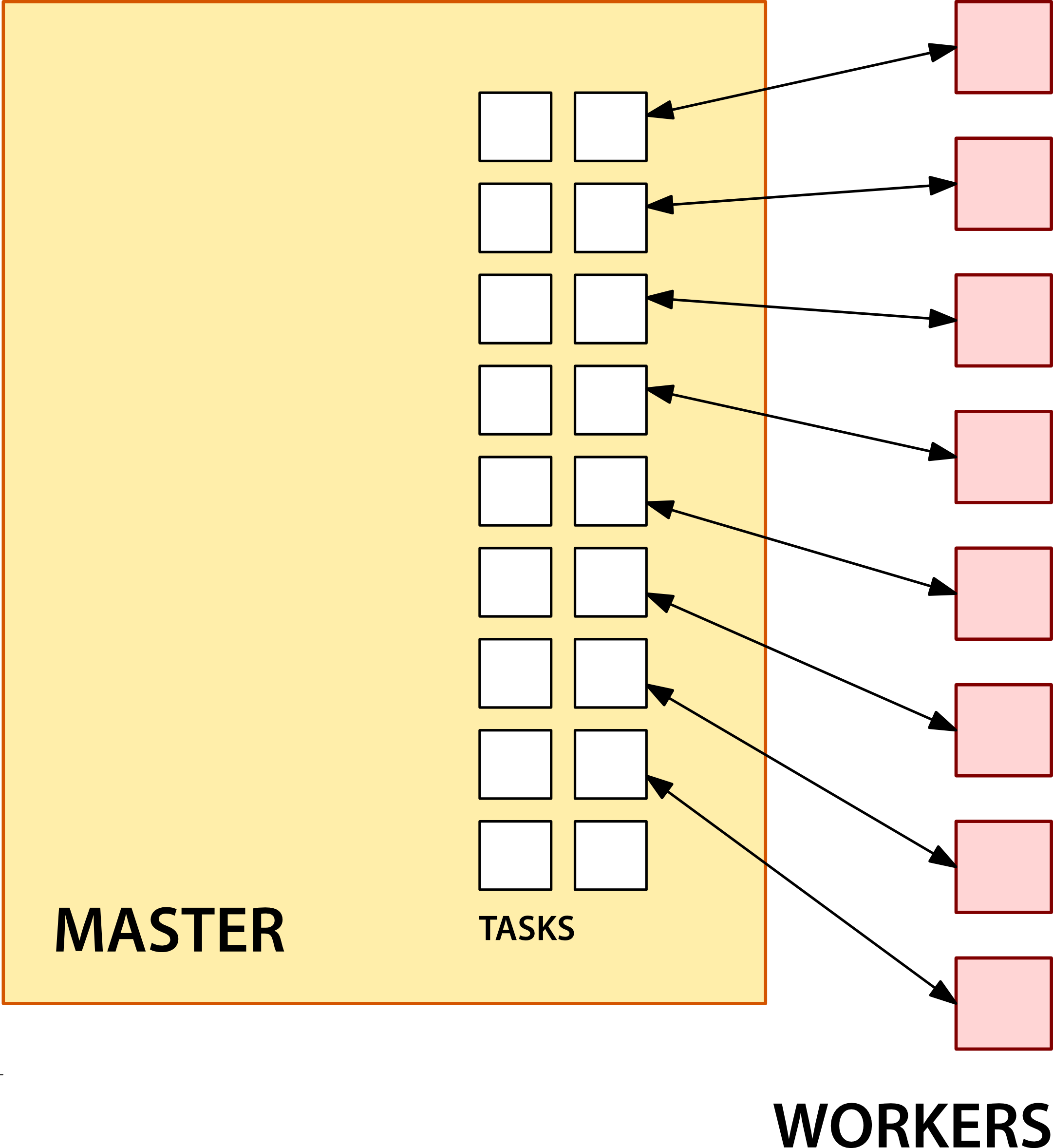}
\caption{
The task farm model. The master node creates and manages a task pool. Each 
worker node takes a task from the pool, returns the result to the master 
and takes the next task, if available. Computations in the pool are 
finished, when all tasks are processed.
}
\label{farm}
\end{minipage}
\begin{minipage}[t]{0.49\textwidth}
\centering
\includegraphics[width=0.9\textwidth]{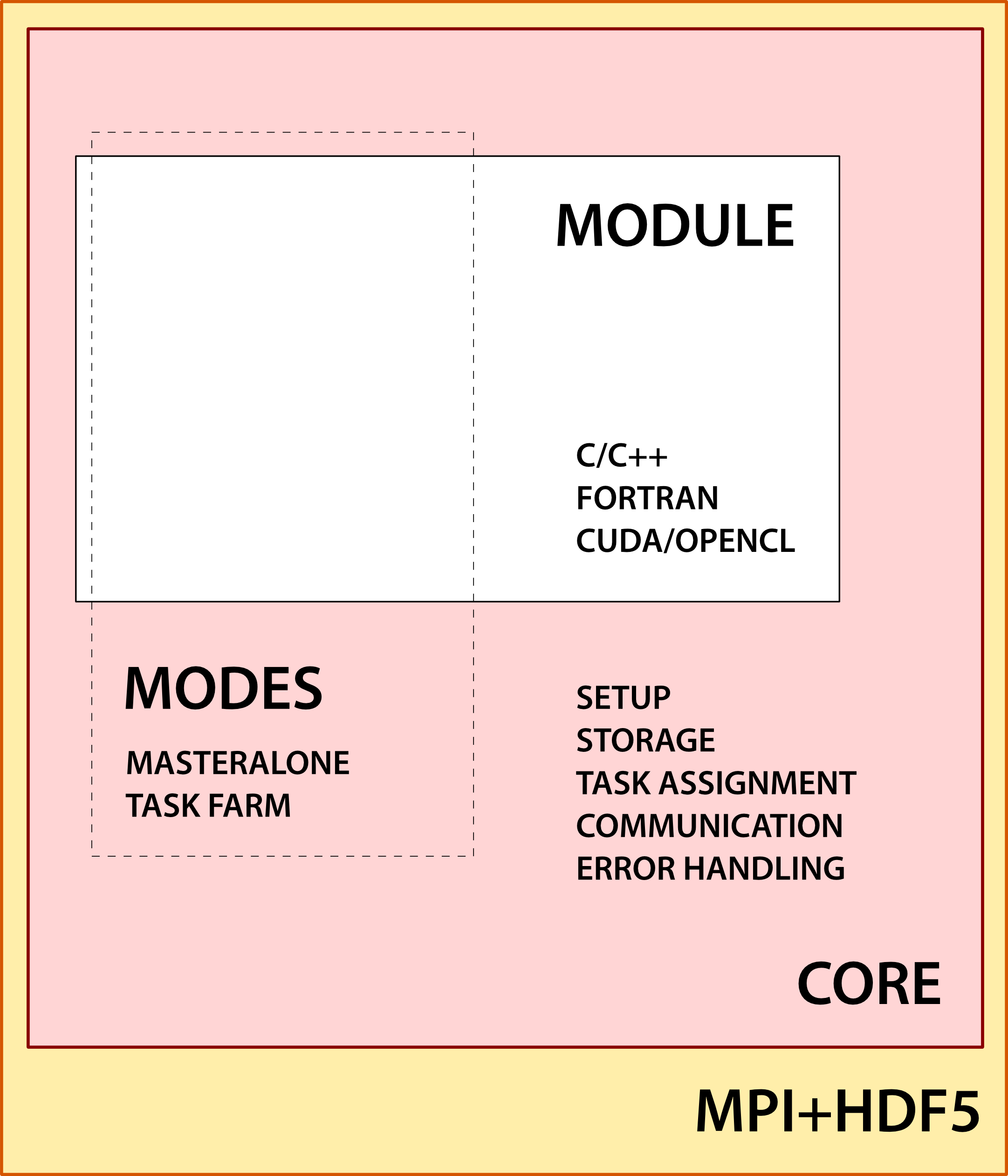}
\caption{
Design of \mechanic{}. An external, user-supplied software 
communicates with the {\em core} through provided API ({\em module}). 
A {\em module} is a C-interoperable code compiled 
in the form of shared library.
}
\label{design}
\end{minipage}
\end{figure}

\noindent
\mechanic{} distributes tasks within the well-known {task farm} (TF) 
communication pattern, which introduces the {master --- worker} 
relationship between nodes (CPUs) in a computing pool (Fig. \ref{farm}). 
The architecture of our software mimics the Unix system architecture in 
terms of the {core---module} scheme. The {core} of the \mechanic{} handles 
the MPI communication, basic setup, task pool configuration and assignment 
(Fig. \ref{design}). A particular numerical problem, implemented in the 
user--supplied {module}, communicates with the {core} through  provided 
Application Programming Interface (API), which makes it possible to manage 
the computations basically in any detail. It applies to any C-interoperable 
programming languages (such as C++, Fortran 2003+, Nvidia CUDA or OpenCL 
frameworks) and allows to reuse existing serial software without a need of 
extensive modifications. The \mechanic{} {core} may be installed 
system-wide and used under common queue control systems. By design, it 
should be possible to run it  under control of any UNIX-like system (Linux 
and Mac OS are actively maintained) and it works uniformly in  single-- and 
multi--CPU environments. The framework may be used in two basic modes: the 
{task farm} (as the default) and a {master-alone} mode (computations are 
performed only on the master node).

One of the key features of \mechanic{} is a dedicated and unified data 
storage model built on the top of the HDF5 standard \cite{hdf}. By default, 
no data management is performed on worker nodes. Instead, the 
user-specified results are sent by these nodes, received by the master node 
and stored in { \em one} global datafile. This helps to reduce the overall 
data processing time and to avoid check-pointing issues. The flexible HDF5 
storage standard makes it possible to use computation results independently 
of the host software, thanks to numerous HDF5--oriented helper applications 
and script tools.

\begin{figure}[!ht]
\vskip0.5\baselineskip
\begin{minipage}[b]{0.49\textwidth}
\centering
\includegraphics[width=0.9\textwidth]{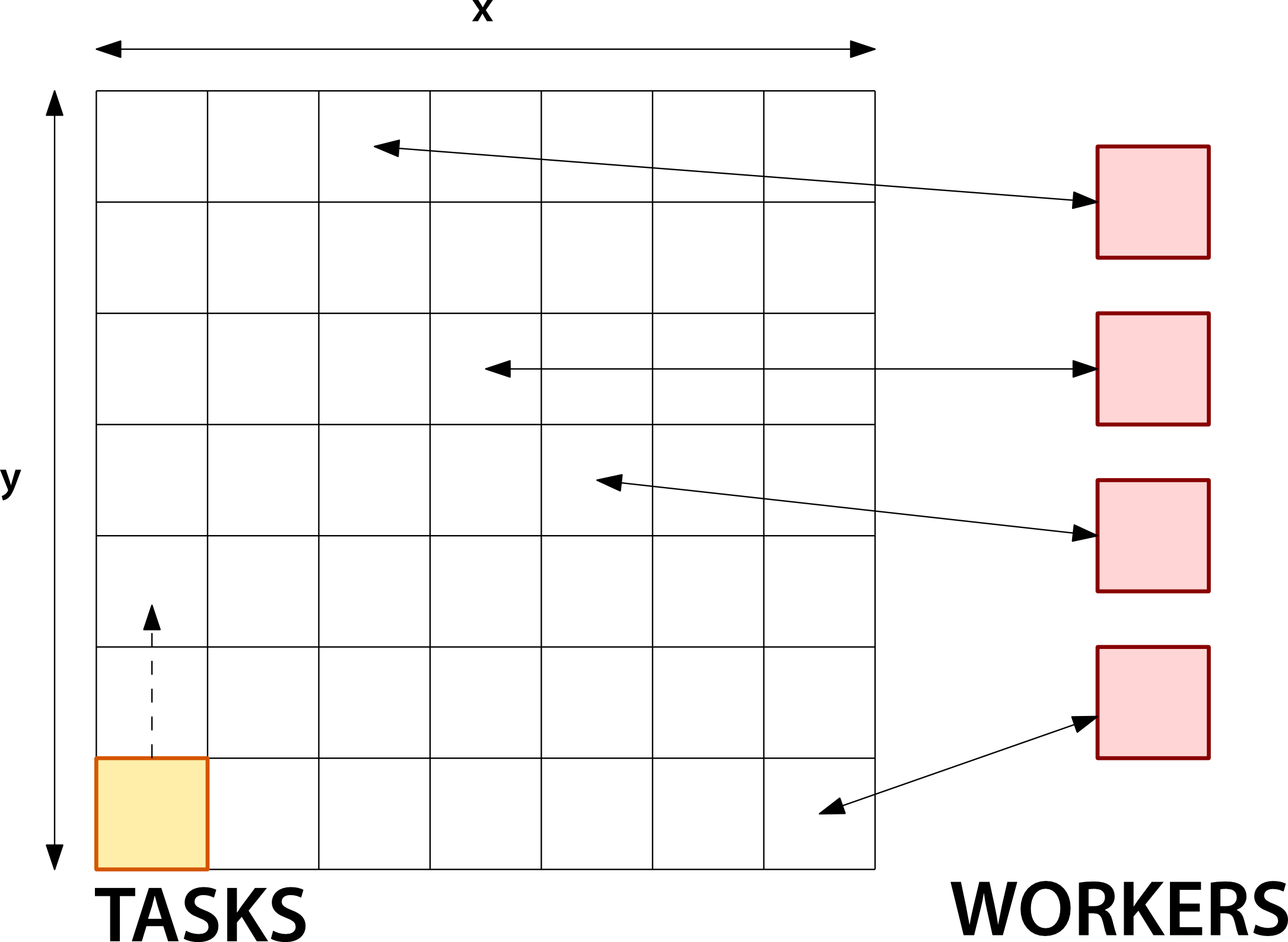}
\caption{
Default task assignment has the structure two-dimensional grid, which  
suits computation of the dynamical map. Each pixel is a standalone 
numerical task, handled by the worker node.
}
\label{map}
\end{minipage}
\begin{minipage}[b]{0.49\textwidth}
\centering
\begin{minipage}[t]{0.9\textwidth}
\begin{verbatim}
module_task_process(task) {

  xstep = (task->xmax - task->xmin)/
          task->xres;
  ystep = (task->ymax - task->ymin)/
          task->yres;

  x = task->xmin + task->x*xstep;
  y = task->ymin + task->y*ystep;

  task->result = your_task_code(x,y);
}
\end{verbatim}
\end{minipage}
\caption{
A sample task preparation (C-pseudocode). Since all information on the 
task is available to the worker node, it may be used to define the initial 
state for each run.
}
\label{task}
\end{minipage}
\end{figure}

The task assignment has a default structure of two-dimensional grid (Fig. 
\ref {map}). The master node creates a pool of computational tasks. Each 
worker node receives a task from the pool, does the computations, returns 
the result to the master node,  and takes the next task to run, if 
available. Coordinates of a specific task are used to define the initial 
state of single--run (Fig. \ref{task}). For instance, to compute a 
dynamical map of a planetary system, we create a mesh grid of initial 
conditions, which are usually constrained through observations. Each 
initial condition is mapped then to a standalone numerical task processed 
by worker nodes. The grid scheme is governed through the API (Fig.~\ref
{task}).


The \mechanic{} has been already extensively tested and used in different 
computing environments, such as large CPU-clusters, workstations or 
laptops. The code remains in early stages of its development, however, it 
provides basic core-functionality, and is suitable to host many 
computational problems of the dynamical astronomy. In principle and by the 
design, the MPI and HDF5 layers are hidden (or rather transparent) to user, 
hence only a very basic knowledge of parallel programming is necessary. 
Advanced topics, such as an additional communication between worker nodes 
or an implementation of different storage layout require parallel computing 
experience. For details of using and developing the \mechanic{} framework 
and its modules, we refer to the user-guide available through the project 
website \cite{mechanic-online}. Sample numerical modules are available 
separately \cite{modules}.

\section{An example of \mechanic{} application: the Arnold web}\label{arnoldweb}
\noindent
To illustrate a simple but non-trivial application of \mechanic, we 
consider  the dynamical system investigated by Froeschl\'e et al.~\cite{science}:
\begin{displaymath}
{\cal H}_{\epsilon}(I_1,I_2,I_3,\phi_1,\phi_2,\phi_3) = 
\frac{1}{2} I_1^2 + \frac{1}{2} I_2^2 + I_3 + 
\frac{\epsilon}{\cos \phi_1+\cos \phi_2+\cos \phi_3+4},
\end{displaymath}
where actions $I_1, I_2, I_3 \in R$ and angles $\phi_1, \phi_2, \phi_3 \in S$
are canonically conjugated phase-space variables, and $\epsilon$ is the 
perturbation parameter. For $\epsilon=0$ the dynamics are integrable. It is 
well known that many qualitative models of the classical mechanics, 
dynamical astronomy, theoretical physics and dynamical systems theory have 
such a general form. It was formulated by Poincar\'e as the very basic 
problem of Celestial Mechanics. According to the Kolmogorov--Arnold--Moser 
theorem (KAM, see e.g. \cite{kam}), if the perturbation is small enough, 
certain non-degeneracy conditions are fulfilled, and if the unperturbed 
motion is {sufficiently non-resonant}, solutions of the perturbed system 
are close to the unperturbed one, and lie on invariant tori. On the same 
torus, all motions are quasi-periodic with the same frequencies. However, 
the KAM theorem does not say much on the dynamics close to the unperturbed 
tori with the {resonant} frequencies.  In the neighbourhood of such a set, 
called {the Arnold web}, the dynamics may be strongly chaotic and complex. 
To study such solutions in detail, basically only numerical methods can be 
applied.
\begin{figure}[!ht]
\centering
\includegraphics[width=0.49\textwidth]{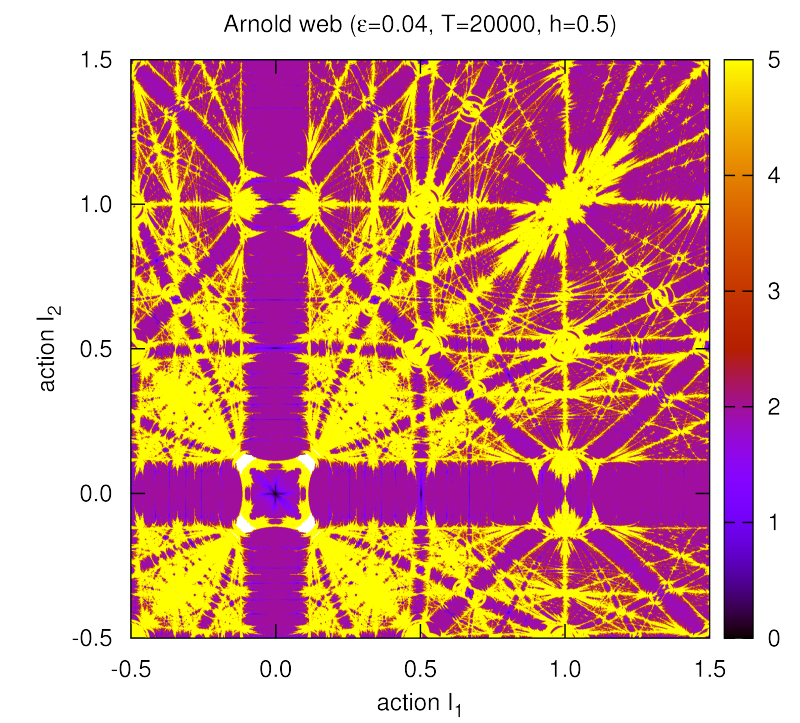}
\includegraphics[width=0.49\textwidth]{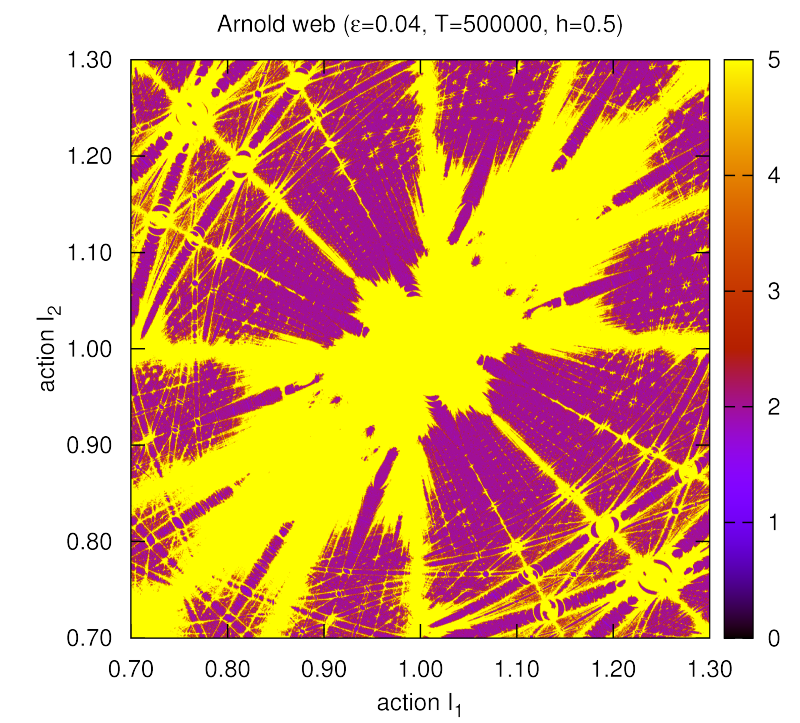}

\includegraphics[width=0.49\textwidth]{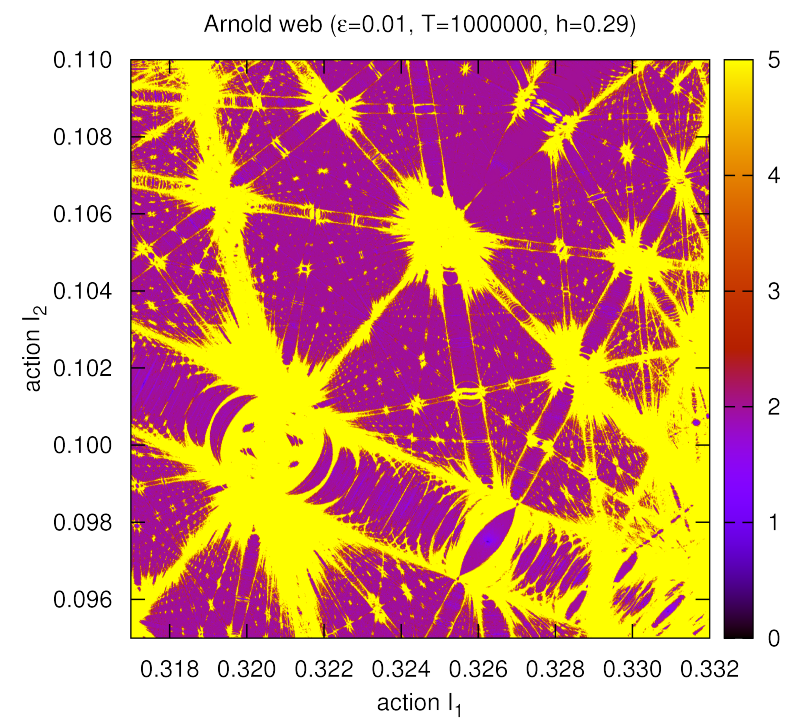}
\includegraphics[width=0.49\textwidth]{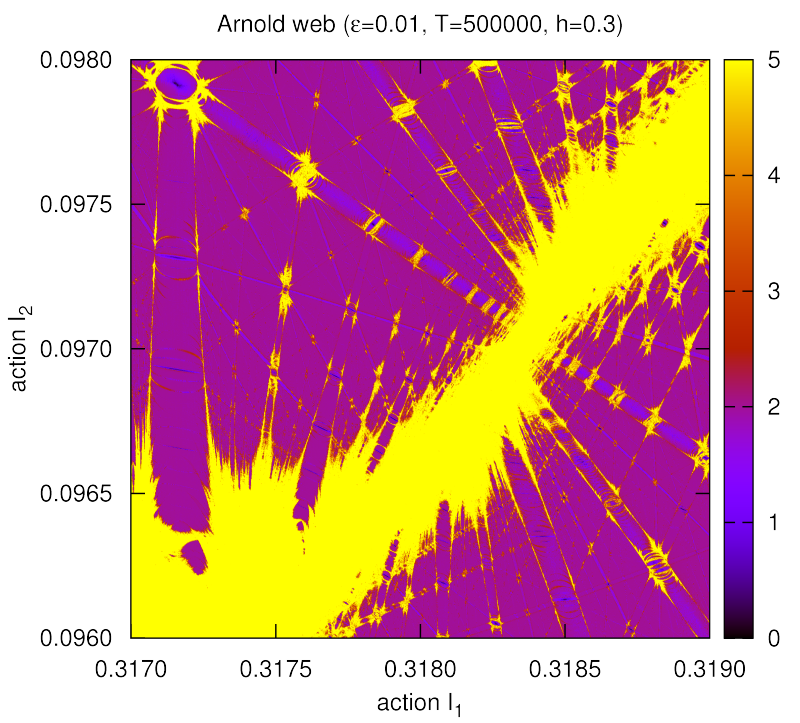}

\caption{
The Arnold web for different values of  perturbation parameter $\epsilon$. 
Subsequent panels are for parts and close-ups of the web for $\epsilon=0.01$
$\epsilon=0.04$. The fast indicator MEGNO has been computed at each point 
of these maps with the help of {\em SABA3} symplectic integration scheme, 
based on the so called tangent map, to distinguish between regular (violet 
regions) and chaotic (yellow regions) motion. Resonances are represented 
and visible as the straight lines. The resolution is 2048 $\times$2048 
pixels (see the text for details). Step-size $h$ and the total integration 
time $T$  are labeled.
}
\label{arnold}
\end{figure}
The structure of the Arnold web can be illustrated at two-dimensional plane 
of actions, $(I_1,I_2)$. Each point of this plane corresponds to 
fundamental frequencies of the unperturbed torus \cite{science}. Resonances 
in this dynamical system are represented by straight lines. Due to infinite 
Fourier expansion of the perturbation, resonances of any order are 
possible. To distinguish between regular and chaotic motions in the 
neighborhood of the resonances, we apply numerical fast indicator, called 
the Mean Exponential Growth factor of Nearby Orbits (MEGNO) invented by 
Cincotta \& Sim\'o \cite{megno}. MEGNO  converges to $\sim2$ for {the 
regular motions}, decreases to 0 for {periodic orbits}, and grows linearly 
with time for {chaotic orbits}. This behaviour can be colour-coded at 
high-resolution dynamical maps, constructed at the frequencies plane (Fig. 
\ref{arnold}) with the help of \mechanic.

To solve the equations of motion and variational equations of the test 
Hamiltonian dynamical system, we applied symplectic integrator {\em SABA}3 
by Laskar \& Robutel \cite{saba} and the tangent map scheme \cite
{mikkola,gozdziewski2008}. Figure \ref{arnold} illustrates the Arnold web 
calculated with very fine resolution of 2048 $\times$ 2048 pixels for two 
values of $\epsilon$, and relatively long integration times ($10^4$--$10^6$
). Subsequent panels are for $\epsilon=0.01$ and $\epsilon=0.04$. The 
close-ups reveal stunning details of the structure of the phase-space. 
Calculations have been run on CPU clusters up to 2048 cores, installed at 
the Pozna\'n Supercomputer Centre ({\tt reef} and {\tt chimera} clusters), 
and took up to a few hours, depending on the integration time and 
step-size. The source code of the numerical module that computes the Arnold 
web, as well as  technical details are available at the project website 
\cite{modules}. An application to  modeling of the radial velocity 
observations, and a study of the global dynamics of the $\nu$~Oct planetary 
system are described in this volume \cite {nuoct}.

\Acknowledgments
\noindent This project is supported by the Polish Ministry of Science and 
Higher Education through grant {\em N/N203/402739}. Computations have been 
conducted within the POWIEW project of the European Regional Development 
Fund in Innovative Economy Programme {\em POIG.02.03.00-00-018/08}.


{\small
\begin{thebibliography}{99}

\bibitem{kam} Arnold, V.I., Weinstein, A., Vogtmann, K., {\em Mathematical Methods of
Classical mechanics}, Springer, 2001
\bibitem{megno} Cincotta, P.~M. \& Sim\'o, C., A\&A, 147, 2000
\bibitem{cincotta2003}
P.~M. {Cincotta}, P.~M., {Giordano}~C.M  \& {Sim{\'o}, C.},
\newblock {\em Phys. D}, 182:151--178, 2003
\bibitem{conway} Conway, B.~A., Chilan, C.~M. \& Wall, B.J.,
Celest. Mech. and Dyn. Astr., 97, 2007
\bibitem{gozdziewski2008}
{Go{\'z}dziewski}, K., {Migaszewski}, C., \& A.~{Musieli{\'n}ski}.
{\em IAU Symposium 249}, 447--460, 2008
\bibitem{gozdziewski2008a}
{Go{\'z}dziewski}, K.,  {Breiter}, S., \& Borczyk, W.,
{\em MNRAS} 383, 989--999, 2008
\bibitem{mpi} Gropp, W. et al. {\em MPI: The Complete Reference}, The MIT Press, 1998
\bibitem{science} Froeschl\'e, C., Guzzo, M. \& Lega, E., Science, 289, 2000
\bibitem{saba} Laskar, J. \& Robutel, P., Celest. Mech., 80, 2001
\bibitem{laskar2010} {{Laskar}, J. \& {Gastineau}, M.}, Nature, 459, 2009
\bibitem{mikkola} {Mikkola}, S. \& {Innanen}, K.,
Celest. Mech. and Dyn. Astr., 74, 1999
\bibitem{nuoct} Go\'zdziewski, K., S\l{}onina, M., K., Rozenkiewicz, A. \& Migaszewski, C., 2012, {\em this volume}
\bibitem{condor} The Condor Research Project, http://www.cs.wisc.edu/condor
\bibitem{hdf} The HDF Group, {\em The HDF5 Standard}, http://www.hdfgroup.org
\bibitem{mechanic-online} The Mechanic User Guide, http://git.astri.umk.pl/projects/mechanic
\bibitem{modules} The Mechanic Modules, http://git.astri.umk.pl/projects/mechanic

\end{thebibliography}
}
\end{document}